\documentclass[sn-mathphys-num]{sn-jnl}

\usepackage{graphicx}%
\usepackage{multirow}%
\usepackage{amsmath,amssymb,amsfonts}%
\usepackage{amsthm}%
\usepackage{mathrsfs}%
\usepackage[title]{appendix}%
\usepackage{xcolor}%
\usepackage{textcomp}%
\usepackage{manyfoot}%
\usepackage{booktabs}%
\usepackage{algorithm}%
\usepackage{algorithmicx}%
\usepackage{algpseudocode}%
\usepackage{listings}%
\usepackage{bm}%
\usepackage{physics}%

\theoremstyle{thmstyleone}%
\theoremstyle{thmstyletwo}%

\theoremstyle{thmstylethree}%

\begin{document}

\title[]{Structure of a single-quantum vortex in $^3$He-A}


\author*[1]{\fnm{Riku} \sur{Rantanen}}\email{riku.s.rantanen@aalto.fi}

\author[1,2]{\fnm{Erkki} \sur{Thuneberg}}

\author[1]{\fnm{Vladimir} \sur{Eltsov}}

\affil[1]{Department of Applied Physics, Aalto University, PO Box 15100, FI-00076 AALTO, Finland}

\affil[2]{\orgdiv{QTF Centre of Excellence}, \orgname{Aalto University}, \orgaddress{\city{Espoo}, \postcode{FI-00076 AALTO}, 
\country{Finland}}}

\abstract{We have performed numerical calculations of the structure of the single-quantum vortex in superfluid $^3$He-A. The GPU-accelerated large-scale numerical simulation is performed in the Ginzburg-Landau model and resolves length scales of both coherence-length-sized hard core and dipolar-length-sized soft core of the vortex. The calculations support previously suggested qualitative structure of the vortex, recently named as eccentric fractional skyrmion, and provide numerical values for the vortex energy, sizes and locations of the hard and soft cores and highly-asymmetric flow profile of the vortex.}




\maketitle

\section{Introduction}\label{sec:introduction}
Quantized vortices are ubiquitous in various superconducting and superfluid systems. They are characterized by their topological properties, such as a fixed winding of the order parameter phase around the vortex core. To accomodate the phase winding in a system with a single-component order parameter, the order parameter amplitude must go to zero at the vortex axis, forming a singular core. The spin-triplet p-wave superfluid $^3$He on the other hand has a multicomponent order parameter, which allows for a variety of more complicated vortex structures.

Vortices in superfluid $^3$He can be non-singular, so that the system remains in the superfluid state even inside the vortex core, which can generally be separated to hard-core and soft-core parts. Inside the hard core of the vortex, the superfluid state differs from the surrounding bulk state. Often the hard core is accompanied by a much larger soft core region, where the order parameter remains in the bulk state, but its orientation changes. The size of the hard core is on the order of the coherence length $\xi \sim 10\,\text{nm}$, while the soft core radius is on the order of the dipole length $\xi_d \sim 10\,\text{µm}$. The large difference in length scales makes it difficult to capture the details of both structures in a single calculation.

Among the superfluid phases of $^3$He, the A phase is unusual as it allows for ``continuous'' vorticity, ie. vortices with no hard core at all. The order parameter of the A phase is
\begin{equation}
    A_{\mu j} = \Delta_A \hat{d}_\mu(\hat{\bm{m}} + i\hat{\bm{n}})_j,
    \label{eq:Aphase}
\end{equation}
where $\Delta_A$ gives the maximum energy gap in the A phase, $\hat{\bm{d}}$ is the spin anisotropy vector, and $\hat{\bm{m}}$ and $\hat{\bm{n}}$ describe the orbital degrees of freedom, so that $\hat{\bm{m}}\times\hat{\bm{n}} = \hat{\bm{l}}$ is the unit vector in the direction of the orbital angular momentum of the Cooper pairs. The order parameter does not have an explicit phase factor $e^{i\varphi}$, as changing the phase by $\varphi$ is equivalent to rotating the $\hat{\bm{m}}$ and $\hat{\bm{n}}$ vectors around $\hat{\bm{l}}$ by the same angle. This gauge-orbit symmetry leads to the important Mermin-Ho relation~\cite{Mermin1976}, that connects the vorticity $\nabla\times\bm{v}_{\mathrm{s}}$ to the orientation of the order parameter:
\begin{equation}
    \bm{\nabla}\times\bm{v}_\mathrm{s} = \frac{\hbar}{4m_3}\sum_{ijk}\epsilon_{ijk}\hat{l}_i(\bm{\nabla}\hat{l}_j\times\bm{\nabla}\hat{l}_k)
    \label{eq:merminho}
\end{equation}
where $\bm{v}_\mathrm{s}$ is the superfluid velocity, and $h/2m_3$ is the quantum of circulation in $^3$He, with $m_3$ being the mass of the $^3$He atom. According to the Mermin-Ho relation, the local vorticity in $^3$He-A is not quantized and can take any value. However, in practical systems with boundaries, the total vorticity is quantized due to the boundaries forcing $\hat{\bm{l}}$ to be perpendicular to the walls~\cite{Ambegaokar1974}, with the number of circulation quanta $\nu$ given by
\begin{equation}
    \nu = \frac{2m_3}{\hbar}\oint \bm{v}_\mathrm{s}\cdot d\bm{r} = \frac{2m_3}{\hbar}\int (\bm{\nabla}\times\bm{v}_\mathrm{s})\cdot d\bm{S},
    \label{eq:circulationquanta}
\end{equation}
where the integrals are over the boundary and the area of the plane perpendicular to the vortex axis, respectively.

The fact that the A phase can carry arbitrary vorticity means that vortices do not necessarily need to have hard cores, as the system can remain in the bulk state everywhere, with only the orientation of the $\hat{\bm{l}}$ vectors varying in space. The Mermin-Ho relation connects the quantization of circulation in the A phase to another topological charge, the skyrmion number. The skyrmion number $N$ for a unit vector $\hat{\bm{l}}$ is defined as
\begin{equation}
    N = \frac{1}{4\pi}\int\hat{\bm{l}}\cdot\left(\frac{\partial\hat{\bm{l}}}{\partial x}\times\frac{\partial\hat{\bm{l}}}{\partial y}\right)dx dy
    \label{eq:skyrmionnumber}
\end{equation}
which describes how many times the $\hat{\bm{l}}$ vector covers the unit sphere when changing across the $x$-$y$ plane. In the vortex, this plane is taken orthogonal to the vortex axis. Combining Eqs.~\eqref{eq:merminho}, \eqref{eq:circulationquanta} and \eqref{eq:skyrmionnumber}, gives
\begin{equation}
    \nu = 2N.
    \label{eq:nu}
\end{equation}

The earliest theoretical works on quantized vortices in the A phase considered a pure phase vortex, with a uniform $\hat{\bm{l}}$ vector and a singular core~\cite{deGennes1973}, but it was quickly found that vorticity stored in the soft-core $\hat{\bm{l}}$ vector texture was energetically preferable~\cite{Mermin1976,Volovik1977,Volovik1981,Maki1983a,Maki1983b,Seppala1983,Fetter1983,Seppala1984,Vulovic1984,Zotos1984}. The first experimental NMR measurements of vortices in the A phase provided evidence for the existence of continuous vortices with no hard core~\cite{Hakonen1982,Hakonen1983}. These structures were identified as the continuous double-quantum vortex with broken axisymmetry (DQV)~\cite{Seppala1983,Fetter1983,Seppala1984,Vulovic1984,Zotos1984,Fetter1987,Salomaa1987,Parts1995,Ruutu1996,Blaauwgeers2000}. Theoretical studies suggested that the single-quantum vortex with a hard core (SQV) is energetically favored, but the observed NMR signatures did not match the predictions~\cite{Seppala1983,Maki1983a,Maki1983b,Hakonen1983,Fetter1983,Vulovic1984,Zotos1984,Fetter1987}. This is explained by SQVs having a larger critical velocity for nucleation, which means that only DQVs are created when rotation is started in the superfluid state~\cite{Blaauwgeers2000}. SQVs were later observed experimentally~\cite{Parts1995}, along with some fraction of DQVs~\cite{Parts1996,Ruutu1996,Parts2000}, when the system was cooled down through the superfluid transition temperature $T_{\mathrm{c}}$ in rotation.

The description of the singly-quantized vortex with both a soft and a hard core has evolved over the years. Although the details of the models differ slightly, they commonly consist of a uniform bulk far from the core (with, for example, $\hat{\bm{l}}=\hat{\bm{x}}$), and a hard core surrounded by a disgyration texture~\cite{Volovik1981,Seppala1983,Fetter1983,Vulovic1984,Fetter1987,Salomaa1987}. Inside the hard core, the superfluid converts to the polar phase~\cite{deGennes1973,Fishman1976,Muzikar1978}. The main differences in the models are the orientation of the disgyration, i.e. whether $\hat{\bm{l}}$ rotates in the $x$-$y$ plane or the $y$-$z$ plane around the hard core, and how the disgyration texture is connected to the surrounding bulk. A structure with a radial disgyration tilted by an angle $\eta$ ($0<\eta<\pi/2$) from the $x$-$y$ plane was suggested in Ref.\ \cite{Karimaki1999}. 

\begin{figure}
    \centering
    \includegraphics[width=\linewidth]{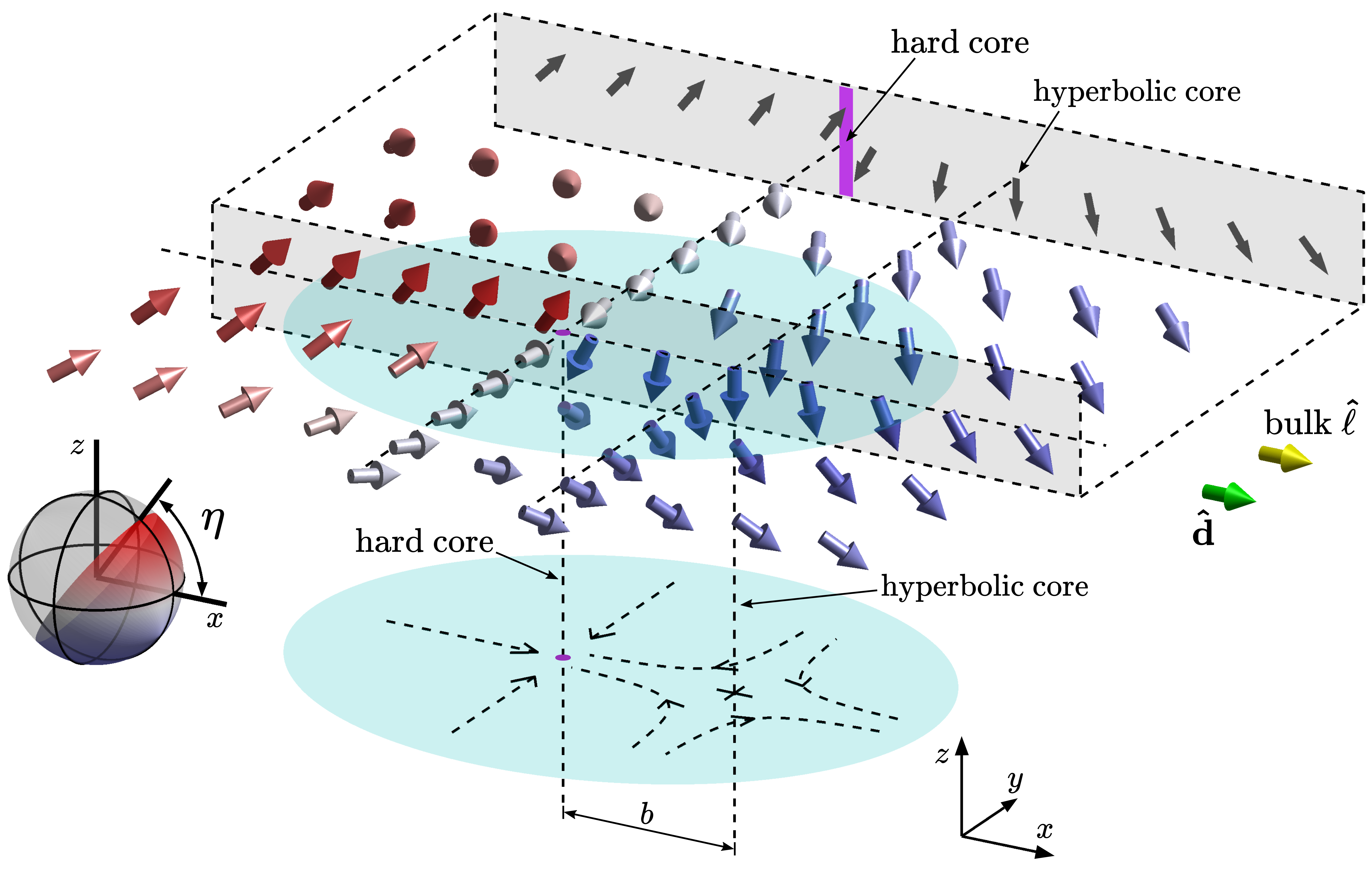}
    \caption{The structure of the single-quantum vortex in superfluid $^3$He-A. Calculations are performed at $p=30\text{ bar}$ and $T = 0.9T_{\mathrm{c}}$ with boundary conditions $\hat{\bm{l}} = \hat{\bm{d}} = \hat{\bm{x}}$. The thick arrows show the $\hat{\bm{l}}$ vector texture, with red color for $\hat{l}_z > 0$, and blue for $\hat{l}_z < 0$. The thin dashed arrows on the bottom illustrate the hyperbolic soft core and the tilted radial disgyration around the hard core, with a distance $b \approx 67\xi \approx 0.25\xi_d \approx 2.3\text{ µm}$ between them. The $\hat{\bm{l}}$ vectors on the $x$ axis are projected onto the $x$-$z$ plane in the top right, showing the rotation of the $\hat{\bm{l}}$ vectors and the hyperbolic soft core with $\hat{\bm{l}}=-\hat{\bm{z}}$. The size of the hard core, marked with purple, is to scale and determined from the calculation as the region where the order parameter amplitude deviates more than five percent from the bulk amplitude. The coverage of the $\hat{\bm{l}}$ vectors in the full SQV texture is shown on the unit sphere in the bottom left, with the tilt angle $\eta \approx 55.4^\circ$.}
    \label{fig:SQVtexture}
\end{figure}

Numerical calculations of the vortex structures in the A phase have typically been limited to continuous structures~\cite{Parts1995,Karimaki1999,Rantanen2023} due to the difficulty in combining the length scales of the hard and soft cores. We present here a numerical calculation of the SQV structure in the Ginzburg-Landau formalism, which for the first time includes simultaneously the coherence-length-sized hard core and the dipole-length-scale soft core. This allows to determine quantitatively the full structure of the vortex, including the locations of hard and soft cores and the angle of the tilted radial disgyration, Fig.~\ref{fig:SQVtexture}.


Structures similar to the SQV in $^3$He-A have been theoretically predicted and experimentally observed in ferromagnetic spin-1 Bose-Einstein condensates~\cite{Takeuchi2022,Huh2024,Lovegrove2016,Sadler2006}. The multicomponent order parameter in these systems allows for a similar nonsingular hard core consisting of a different condensate phase from the surrounding bulk. Just like in the SQV, the spin texture (analogous to the $\hat{\bm{l}}$ vectors) rotates to cover one half of the unit sphere, with a hard-core defect in the texture. If the hard core is shifted off center like in the SQV, the structure is called an eccentric fractional skyrmion~\cite{Takeuchi2022,Huh2024}, due to the half-integer skyrmion number $N$. Unlike in $^3$He-A, where the DQV appears to be stable, the disassociation of a continuous skyrmion texture into two hard-core fractional skyrmions has also been observed in spinor BECs~\cite{Weiss2019,Huh2024}.

\section{Numerical simulation}\label{sec:numerics}

We calculate the structure of the SQV using the Ginzburg-Landau (GL) formalism. For details of the theory and the coefficients used, see Ref.~\cite{Rantanen2024}. The superfluid state is described by the $3\times 3$ order parameter matrix $A_{\mu j}(\bm{r})$, where the first index corresponds to spin degrees of freedom and the second to orbital ones. We calculate the GL free energy $\mathcal{F}$ as a functional of $A_{\mu j}(\bm{r})$ and minimize it to find the lowest-energy order parameter state.

The geometry of the system is discretized as a mesh composed of tetrahedral elements. The order parameter matrix is defined at each point in the mesh, i.e. at the vertices of the tetrahedrons. To calculate the energy density inside a tetrahedron we use linear interpolation and barycentric coordinates. The energy of a single element can then be calculated by integrating the energy density over the volume of the tetrahedron, and the total energy of the system is the sum of the energies of all elements in the mesh.

For a tetrahedron with four vertices $\bm{r}_1$, $\bm{r}_2$, $\bm{r}_3$ and $\bm{r}_4$, the barycentric coordinates $\lambda_i(\bm{x})$ are defined so that $\lambda_i(\bm{r}_j) = \delta_{ij}$, and $\lambda_1+\lambda_2+\lambda_3+\lambda_4 = 1$, i.e. each coordinate $\lambda_i$ is a linear function inside the element so that its value is 1 at the corresponding vertex and 0 on the opposite face. This coordinate system is useful as it allows easy linear interpolation in 3D space. The order parameter matrix $A_{\mu j}$ inside a tetrahedron is approximated as
\begin{equation}
    A_{\mu j}(\bm{x}) = \sum_{i=1}^4 \lambda_i(\bm{x})A_{\mu j}^i\,,
    \label{eq:bccinterpolation}
\end{equation}
where $A_{\mu j}^i$ is the value of the order parameter defined at vertex $i$ of the tetrahedron. The order parameter matrices are given as 18 real values, 9 each for the real and imaginary components so that
\begin{equation}
    A_{\mu j}^i = a_{\mu j i} + ib_{\mu j i}.
    \label{eq:complexA}
\end{equation} 
The gradients of the order parameter inside the tetrahedron are given by
\begin{equation}
    \frac{\partial A_{\mu j}}{\partial x_l} = \sum_{i=1}^4 \frac{\partial A_{\mu j}}{\partial \lambda_i}\frac{\partial \lambda_i}{\partial x_l} = \sum_{i=1}^4 A_{\mu j}^i \frac{\partial \lambda_i}{\partial x_l}\,,
    \label{eq:bccgradients1}
\end{equation}
and the gradients of the barycentric coordinates are
\begin{equation}
    \bm{\nabla}\lambda_i = -\frac{S_i}{3V}\hat{\bm{n}}^i\,,
    \label{eq:bccgradients2}
\end{equation}
where $S_i$ is the area of the face opposite vertex $i$, $\hat{\bm{n}}^i$ is the outward pointing normal of that face, and $V$ is the volume of the tetrahedron. The barycentric gradients are constant inside the element.

The linear approximation has the additional benefit of simplifying the integration considerably. For example, the first fourth-order bulk energy term inside a tetrahedron is calculated as
\begin{align}
    \int_V \beta_1 \left\vert\Tr{A^TA}\right\vert^2 dV &= \int_V \beta_1\left\vert\sum_{i,j=1}^4 \lambda_i\lambda_j \Tr{(A^i)^T A^j}\right\vert^2 dV \nonumber \\
    &= \beta_1 \int_V \sum_{i,j,k,l=1}^4 \lambda_i\lambda_j\lambda_k\lambda_l \Tr{(A^i)^T A^j}\Tr{(A^k)^T A^l}^* dV \nonumber \\
    &= \beta_1 \sum_{i,j,k,l=1}^4 \Tr{(A^i)^T A^j}\Tr{(A^k)^T A^l}^* \int_V \lambda_i\lambda_j\lambda_k\lambda_l dV.
    \label{eq:beta1integral}
\end{align}
The integration adds a constant multiplier to each term in the sum that depends only on the volume of the tetrahedron and the indices $i$, $j$, $k$ and $l$. These integrals can be precomputed for the whole mesh, which saves time during the energy minimization. The values of the second-order integrals of the barycentric coordinates are
\begin{equation}
    \int_V \lambda_i\lambda_j dV = \begin{cases}
        V/10, &\text{ when } i=j \\
        V/20, &\text{ when } i\neq j
    \end{cases}
    \label{eq:bccint1}
\end{equation}
and the fourth-order integrals are
\begin{equation}
    \int_V\lambda_i\lambda_j\lambda_k\lambda_l dV = \begin{cases}
        V/35, &\text{ when } i=j=k=l  \\
        V/140, &\text{ when three indices are equal} \\
        V/210, &\text{ when two pairs of indices are equal} \\
        V/420, &\text{ when only two indices are equal }\\
        V/840, &\text{ when no indices are equal}
    \end{cases}
    \label{eq:bccint2}
\end{equation}

The integral of the gradient energy densities is similarly simplified, for example the $K_1$ term becomes
\begin{align}
    \int_V K_1 \nabla_k A^*_{\alpha j} \nabla_k A_{\alpha j} dV &= K_1\int_V\sum_{\alpha,j,k=1}^3 \frac{\partial A^*_{\alpha j}}{\partial x_k}\frac{\partial A_{\alpha j}}{\partial x_k}dV \nonumber \\
    &= K_1\int_V\sum_{\alpha,j,k=1}^3\left[\left(\sum_{m=1}^4 A_{\alpha j}^{m*}\frac{\partial \lambda_m}{\partial x_k}\right)\left(\sum_{n=1}^4 A_{\alpha j}^n\frac{\partial \lambda_n}{\partial x_k}\right)\right]dV \nonumber \\
    &= K_1 \sum_{m,n=1}^4 \left[\left(\sum_{k=1}^3 \frac{\partial \lambda_m}{\partial x_k}\frac{\partial \lambda_n}{\partial x_k}\right) \Tr{(A^m)^\dag A^n}\right]\int_V dV.
    \label{eq:K1}
\end{align}
The integral in Eq.~\eqref{eq:K1} only adds a constant factor equal to the volume of the tetrahedron.

In order to perform efficient numerical minimization of the total energy, we utilize the functional gradients, i.e. the derivatives of the energy with respect to the input variables, the order parameter values at the vertices. The functional gradient of the bulk energy $f_{\text{bulk}}$ with respect to the real part $a_{\mu j k}$ is
\begin{align}
    \frac{\partial f_{\text{bulk}}}{\partial a_{\mu j k}} &= \frac{\partial f_{\text{bulk}}}{\partial A_{\mu j}}\frac{\partial A_{\mu j}}{\partial a_{\mu j k}} + \frac{\partial f_{\text{bulk}}}{\partial A_{\mu j}^*}\frac{\partial A_{\mu j}^*}{\partial a_{\mu j k}} \nonumber \\
    &= \left[\frac{\partial f_{\text{bulk}}}{\partial A_{\mu j}^*} + \left(\frac{\partial f_{\text{bulk}}}{\partial A_{\mu j}^*}\right)^*\right]\lambda_k \nonumber
    = 2\lambda_k\Re\frac{\partial f_{\text{bulk}}}{\partial A_{\mu j}^*}
    \label{eq:bulkaijk}
\end{align}
and similarly for the imaginary component $b_{\mu j k}$
\begin{equation}
    \frac{\partial f_{\text{bulk}}}{\partial b_{\mu j k}} = 2\lambda_k\Im\frac{\partial f_{\text{bulk}}}{\partial A_{\mu j}^*}\,,
    \label{eq:bulkbijk}
\end{equation}
where
\begin{align}
    \frac{\partial f_{\text{bulk}}}{\partial A_{\mu j}^*} &= \alpha A_{\mu j} + 2\beta_1 A_{\mu j}^*\Tr{A^T A} + 2\beta_2 A_{\mu j}\Tr{A^\dag A} \nonumber \\ &+ 2\beta_3[A(A^\dag A)^*]_{\mu j} + 2\beta_4(AA^\dag A)_{\mu j} + 2\beta_5(A^*A^TA)_{\mu j}.
    \label{eq:dfbulkdA}
\end{align}
The functional gradients of the gradient energy are found using the linear approximation of Eq.~\eqref{eq:bccgradients1}:
\begin{equation}
    \frac{\partial f_{\text{grad}}}{\partial a_{\mu j k}} = 2\Re\frac{\partial f_{\text{grad}}}{\partial (A_{\mu j}^k)^*}\quad\text{and}\quad \frac{\partial f_{\text{grad}}}{\partial b_{\mu j k}} = 2\Im\frac{\partial f_{\text{grad}}}{\partial (A_{\mu j}^k)^*}\,,
    \label{eq:gradaijkbijk}
\end{equation}
where the derivative of the gradient energy density with respect to the order parameter matrices at the vertices is
\begin{align}
    \frac{\partial f_{\text{grad}}}{\partial (A_{\mu j}^k)^*} &= K_1\sum_{n=1}^4\sum_{m=1}^3\frac{\partial\lambda_n}{\partial x_m}\frac{\partial\lambda_k}{\partial x_m}A_{\mu j}^n \nonumber \\
    &+ K_2\sum_{n=1}^4\sum_{m=1}^3\frac{\partial\lambda_k}{\partial x_j}\frac{\partial\lambda_n}{\partial x_m}A_{\mu m}^n \nonumber \\
    &+ K_3\sum_{n=1}^4\sum_{m=1}^3\frac{\partial\lambda_n}{\partial x_j}\frac{\partial\lambda_k}{\partial x_m}A_{\mu m}^.
    \label{eq:dfgraddA}
\end{align}
Note that with the coefficients used in simulations, $K_1 = K_2 = K_3 = K$.

For the minimization of the free energy, we use the Limited-memory Broyden-Fletcher-Goldfarb-Shanno (L-BFGS) algorithm~\cite{Liu1989}. This algorithm is best suited for problems with a large number of minimization parameters. The method uses a relatively small amount of memory by approximating the Hessian matrix with the gradient information from previous iterations.

The energy and gradient calculation for each individual tetrahedral element can be done in parallel. For this reason, we take advantage of graphics processing units (GPUs), which are designed for such massively parallelizable calculations. The minimization is also performed on the GPU, using an adapted version of the CudaLBFGS library~\cite{CudaLBFGS}.

\section{The single-quantum vortex}\label{sec:sqv}

\begin{figure}
    \centering
    \includegraphics[width=\linewidth]{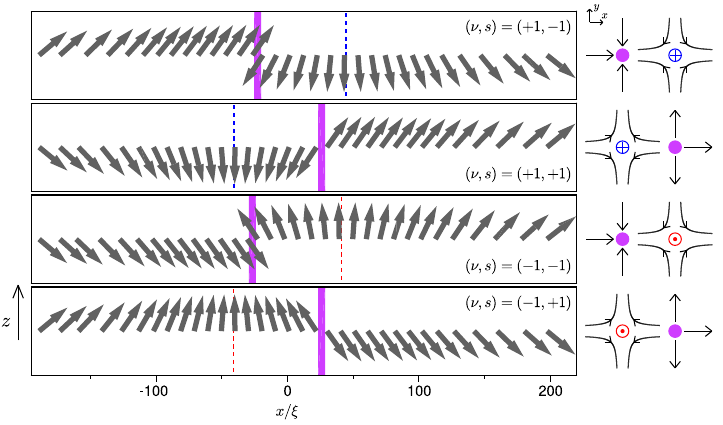}
    \caption{Four types of SQV structures. The left panels show the $x$-$z$ projection of the $\hat{\bm{l}}$ vector texture along the $x$ axis. The four types are labeled by  $(\nu, s)$, with $\nu$ signifying circulation and $s$ the texture symmetry as defined in Eq.~(\ref{eq:ltilt}). The illustrations on the right side show the top down view of the SQV texture on the $x$-$y$ plane. The hard cores are marked with thick bars on the left and solid circles on the right. The center of the hyperbolic soft core is marked with a dashed line on the left and crosses or dots on the right, for $\hat{\bm{l}}=-\hat{\bm{z}}$ and $\hat{\bm{l}}=\hat{\bm{z}}$, respectively.}
    \label{fig:vortextypes}
\end{figure}

In a bulk system the single-quantum vortex is the lowest energy vortex structure in $^3$He-A. The double-quantum vortex, although stable, has a higher energy due to the increased circulation around the vortex. The flow energy outside the core of a vortex with $\nu$ quanta of circulation in the A phase is given by
\begin{equation}
    \mathcal{F}_{\text{kin}} = \int_0^{2\pi}\int_{R_{\mathrm{c}}}^{R_{\text{out}}}\left[\frac{1}{2}\rho_{\mathrm{s}\perp}v_{\mathrm{s}}^2 - \frac{1}{2}(\rho_{\mathrm{s}\perp}-\rho_{\mathrm{s}\parallel})(\hat{\bm{l}}\cdot\bm{v}_{\mathrm{s}})^2\right]rdrd\phi
    \label{eq:fkin1}
\end{equation}
where $R_{\mathrm{c}}$ is the core radius, $R_{\text{out}}$ is some outer cutoff radius, and $\rho_{\mathrm{s}\perp}$ and $\rho_{\mathrm{s}\parallel}$ are the superfluid densities perpendicular and parallel to $\hat{\bm{l}}$, respectively. In the simplest model $\bm{v}_{\mathrm{s}} = (\hbar/2m_ 3)(\nu/r)\hat{\bm{\phi}}$ and  $\hat{\bm{l}}$ is uniform in the bulk, for example $\hat{\bm{l}}=\hat{\bm{x}}$. The total flow energy is
\begin{equation}
    \mathcal{F}_{\text{kin}} = \frac{\pi}{2}(\rho_{\mathrm{s}\perp}+\rho_{\mathrm{s}\parallel})\left(\frac{\hbar}{2m_3}\right)^2 \nu^2 \ln{\left(\frac{R_{\text{out}}}{R_{\mathrm{c}}}\right)}.
    \label{eq:fkin2}
\end{equation}
From Eq.~\eqref{eq:fkin2}, it is clear that the DQV with $\nu=2$ has the flow energy essentially larger than two SQVs with $\nu=1$. The same conclusion is obtained by a more accurate calculation of the asymptotic behavior in Ref.\ \cite{Volovik1981}, although the deviation of $\hat{\bm{l}}$ from a constant is underestimated there.

In contrast to the hard-core vortices found in the B phase, the vorticity of the SQV is all localized in the soft core of the vortex instead of the hard core. The existence of the hard core is necessary to resolve the kink that appears in the $\hat{\bm{l}}$ vector texture when connecting the uniform bulk texture to the soft core which covers one half of the unit sphere, see Eqs.~\eqref{eq:skyrmionnumber} and \eqref{eq:nu}. Due to the anisotropy of the uniform bulk $\hat{\bm{l}} = \hat{\bm{x}}$ texture, the vortex is not axially symmetric, but instead the hard core shifts along the $x$ axis.

To understand the structure of the SQV, it is useful to first consider a simple phase vortex with $\hat{\bm{l}} = \hat{\bm{x}}$ everywhere and $\hat{\bm{m}}+i\hat{\bm{n}} = e^{i\phi}(\hat{\bm{y}}+i\hat{\bm{z}})$. Here $\phi$ is an azimuthal angle in the $x$-$y$ plane, coinsiding in this simple model with the phase of the order parameter. At the vortex axis, the phase $\phi$ is not well defined and the only solution is to suppress the superfluidity completely in the core. However, this suppression costs a lot of condensation energy. The energy of the state can be lowered by deviating the $\hat{\bm{l}}$ vector from the bulk texture to a radial disgyration close to the hard core, so that
\begin{equation}
    \hat{\bm{l}} = s\hat{\bm{r}},\quad \hat{\bm{m}}+i\hat{\bm{n}} = \hat{\bm{z}} + is\hat{\bm{\phi}},
    \label{eq:disgyration}
\end{equation}
where $s=\pm1$ differentiates between outward or inward radial disgyration. For a radial disgyration, the superfluidity in the hard core is not fully suppressed, but instead converts to the polar phase where the $\hat{\bm{n}}$ term in the order parameter vanishes and $\hat{\bm{m}} = \hat{\bm{z}}$~\cite{Fishman1976,Muzikar1978}. Connecting the radial $\hat{\bm{l}}$ texture close to the hard core to the uniform $\hat{\bm{l}}=\hat{\bm{x}}$ bulk makes the core asymmetric. On one side of the vortex the radial $\hat{\bm{l}}$ smoothly connects to the bulk texture, while on the opposite side $\hat{\bm{l}}$ has to rotate $180^\circ$. The gradient energy stemming from the $\hat{\bm{l}}$ rotation can be reduced by shifting the hard core to the $\pm x$ direction, depending on the value of $s$. To accomodate the core shift, the radial disgyration (and the orientation of the polar core) tilts, so that
\begin{equation}
    \hat{\bm{l}} = s\left[\hat{\bm{y}}\sin\phi + \cos\phi(\hat{\bm{x}}\cos\eta + \hat{\bm{z}}\nu\sin\eta)\right]
    \label{eq:ltilt}
\end{equation}
where $\nu = 1$ for positive circulation along $\hat{\bm{z}}$ and $\nu=-1$ for negative circulation, i.e. for the antivortex. The angle $\eta$ is constant and depends on the full texture of the SQV, including the hard core. We follow the notation of Ref.~\cite{Karimaki1999} and define $\eta$ as the maximum angle between $\hat{\bm{l}}$ and the $x$ axis at the disgyration. From Eq.~\eqref{eq:ltilt} it is clear that there are four possible structures of a vortex aligned with $z$ axis depending on the values of $\nu$ and $s$. These structures are shown in Figure~\ref{fig:vortextypes}.

The singular vortex has symmetry $m_y'$, which is often called $v$ in the context of vortices of superfluid $^3$He. Here $m_y$ denotes reflection in the plane perpendicular to the $y$ axis and the prime denotes time inversion. In the singular vortex the symmetries $2_y'$ and $\overline1$ are spontaneously broken. Here $2_y'$ is two-fold rotation around $y$ combined with time inversion, and the inversion $\overline1$ is the product of the previous operators. The two degenerate forms of SQV with the same circulation are obtained from each other either by $2_y'$ or by $\overline1$. The opposite-circulation states are degenerate with these only in the absence of rotation.

\section{Results}\label{sec:results}

\begin{figure}
    \centering
    \includegraphics[width=\linewidth]{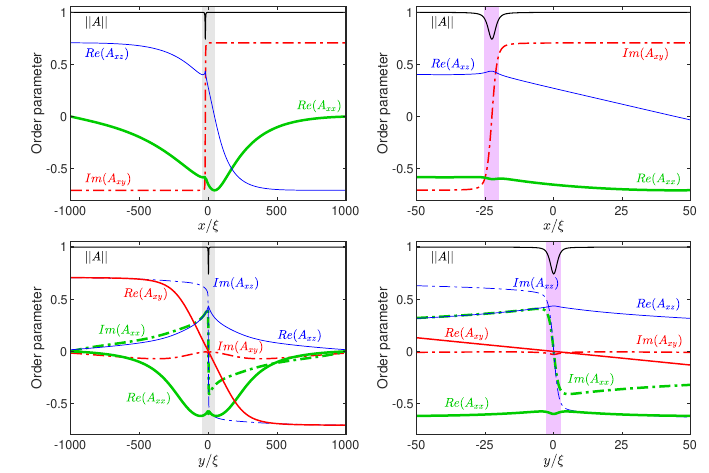}
    \caption{The order parameter components $A_{\mu j}$ on the $x$ axis (top panels) and on the $y$-direction axis through the hard core (bottom panels). The left panels display the full computational domain, where the shaded regions are zoomed in in the right panels. The purple areas in the right panels show the hard core region where the amplitude drops below 0.95 of the bulk value. Real components are marked with solid lines and imaginary components with a dash-dotted lines. The $A_{xx}$ components are marked with thick green lines, the $A_{xy}$ components with red lines and the $A_{xz}$ components with thin blue lines. The order parameter amplitude $||A|| = \sqrt{\Tr{A^\dag A}}$ is shown with a solid black line at the top of each panel. All values are normalized to the bulk order parameter amplitude $||A||_{\text{bulk}} = \sqrt{2}\Delta_A$.}
    \label{fig:orderparameter}
\end{figure}
\begin{figure}
    \centering
    \includegraphics[width=\linewidth]{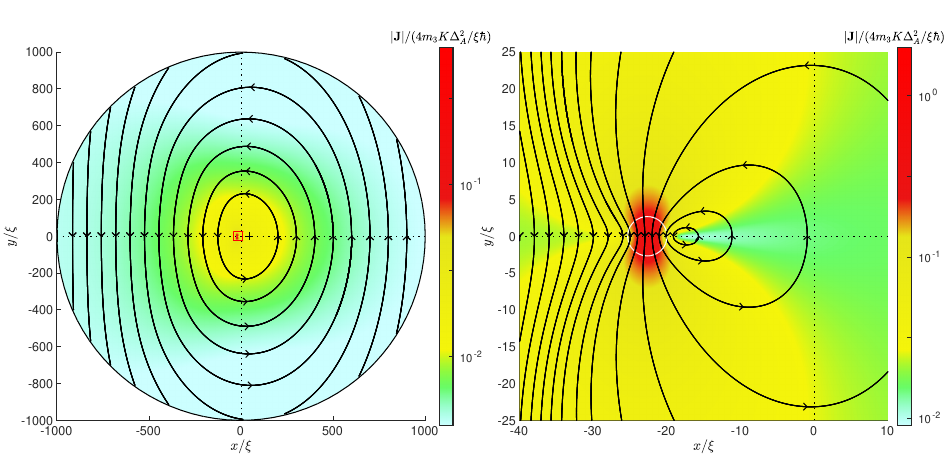}
    \caption{The superfluid mass currents in the SQV. The left panel shows the current density for the whole system, with lines representing streamlines, i.e. the direction of the flow. The density of the streamlines does not correspond to the strength of the flow. The magnitude of the flow is indicated by colors in the units of $4m_3K\Delta_A^2/\xi\hbar$. The right panel shows a zoomed in section marked by the small square in the left panel. In the right panel, the hard core with polar phase is marked with a white circle. The highest current density  is localized inside the hard core. Note also how the center of the circulation is shifted right from the hard core.}
    \label{fig:currents}
\end{figure}
\begin{figure}
    \centering
    \includegraphics[width=0.7\linewidth]{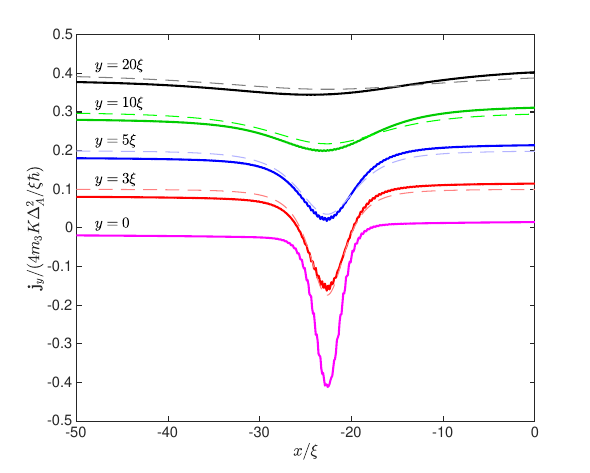}
    \caption{The $y$ component of the superfluid mass current as a function of $x$ plotted at $y = 0$, $y = 3\xi$, $y = 5\xi$, $y = 10\xi$ and $y = 20\xi$. For clarity, the curves are shifted relative to each other on the vertical axis by $0.1$. The solid lines are values from the numerical calculation, and dashed lines represent the model in Eq.~\eqref{e.jcdens}, evaluated without fitting parameters with the tilt $\eta = 55.39^\circ$ and with the location of the hard core $x = -22.67\xi$ as in the structure in Fig.~\ref{fig:SQVtexture}. The asymmetry in the current is due to the superfluid flow around the vortex, which is not included in the model. The magnitude of the current increases near the hard core in approaching the $x$ axis, showing the focusing effect of the hard core.}
    \label{fig:xcurrents}
\end{figure}

We calculate the full structure of the SQV by minimizing the GL free energy as described in Section~\ref{sec:numerics}. The geometry is a disk with a radius of $R_{\text{calc}} = 1000\xi$, with two layers of points in the $z$ direction with a distance of $\xi$. The two layers are set with periodic boundary conditions to simulate an infinitely long vortex in the $z$ direction. On the radial boundary at $r = 1000\xi$, we set the order parameter to have a constant bulk value of $A_{\mu j} = \Delta_A\hat{x}_\mu e^{i\nu\phi}(i\hat{\bm{y}}-\hat{\bm{z}})_j$ where $\Delta_A$ is the bulk A phase gap. The distance between points in the $x$-$y$ plane is $\xi/3$ in the region $r < 100\xi$, and then gradually increases to $\sim 3\xi$ at $r = 1000\xi$. In total the mesh has approximately $4.9$ million points and $14.6$ million tetrahedral elements.

The calculations are done at a pressure of $p = 30\text{ bar}$ and temperature $T = 0.9T_{\mathrm{c}}\approx 2.2\text{ mK}$, with a magnetic field of $H = 100\text{ G}$ oriented along the $z$ axis. The temperature and pressure control the $\beta$ parameters of the GL theory and the ratio $\xi/\xi_d$ \cite{Rantanen2024}. The initial order parameter state given to the minimization is a uniform bulk $\hat{\bm{l}} = \hat{\bm{x}}$ with a $2\pi\nu$ phase winding in order to find a vortex state with a single quantum of circulation. To speed up the calculation, no external rotation is applied, as it is not expected to have an effect on the length scales of a single vortex.

We simulate the four types of SQV structures shown in Figure~\ref{fig:vortextypes}. The value of $\nu$ is given as the direction of the phase winding in the initial state and at the constant radial boundary. The two possible $s$ states are found by biasing the initial texture by tilting the $\hat{\bm{l}}$ vectors either towards or away from the $x$ axis near the $y$ axis. We find that the four vortex types have equivalent energies to within $0.04\%$ (with the energy of the uniform bulk A phase removed). The rest of the data in the paper is given for the $(\nu,s) = (+1,-1)$ state.

Figure~\ref{fig:SQVtexture} shows the $\hat{\bm{l}}$ vector texture for the SQV. The hard core of the vortex shifts along the $x$ axis to $x = -22.67\xi$ from the center of the cylinder, and a hyperbolic soft core texture with $\hat{\bm{l}}=-\hat{\bm{z}}$ is formed on the other side at $x = 44.50\xi$. The shift directions are determined by the value of $s$ and the sign of $\hat{\bm{l}}$ in the hyperbolic core is determined by $\nu$, as seen in Figure~\ref{fig:vortextypes}. The radial disgyration texture is tilted from the $x$-$y$ plane as described by Eq.~\eqref{eq:ltilt}, with an angle $\eta\approx 55.4^\circ$. This is reflected in the $\hat{\bm{l}}$ vector coverage of the unit sphere, shown in Figure~\ref{fig:SQVtexture}.

All non-zero order parameter values $A_{\mu j}$ along the $x$ and $y$ axes through the hard core are shown in Figure~\ref{fig:orderparameter}. The asymmetric structure of the SQV is clearly visible. In the hard core, the $\hat{\bm{n}}$ vector vanishes, leaving only the tilted $\hat{\bm{m}}$ vector of the polar phase. Outside the hard core on the $x$ axis, $\hat{\bm{n}}$ stays fixed while $\hat{\bm{m}}$ rotates slowly. The structure on the $y$ axis through the hard core is more complicated, due to the phase winding around the core. The order parameter amplitude is slightly reduced from the bulk value inside the hard core.

The superfluid mass current density can be calculated within the Ginzburg-Landau formalism as~\cite{Thuneberg1987}
\begin{equation}
    {j}_i = \frac{4m_3K}{\hbar}\Im\left(A_{\mu i}^*\frac{\partial A_{\mu j}}{\partial x_j} + A_{\mu j}^*\frac{\partial A_{\mu j}}{\partial x_i} + A_{\mu j}^*\frac{\partial A_{\mu i}}{\partial x_j}\right),
\end{equation}
and it is plotted in Figure~\ref{fig:currents}. Due to the asymmetric structure of the SQV, the flow pattern is highly asymmetric as well with the streamlines forming elliptical shapes that are flattened on the left-hand side. Interestingly, the center of circulation is not in the hard core or even at the hyperbolic soft core, but instead between the two, next to the polar core. The flow on the $-x$ side of the circulation center is strongly concentrated through the polar core. The strong flow through the core in the $-y$ direction is compensated on the $+x$ side of the circulation center in a very large area.

It looks like the polar core acts as a channel for the mass flow in the $-y$ direction. This can be understood using the following model. 
We write the order parameter in the neighborhood of the hard core as
\begin{eqnarray}
A_{\mu j}(x,y)=\Delta_A\hat{d}_\mu(\hat{\bm m}+if\hat{\bm n})_j
\label{e.opmhca}\end{eqnarray}
with
\begin{eqnarray}
\hat{\bm l}&=&-\hat{\bm y}\sin\phi-\cos\phi(\hat{\bm x}\cos\eta+\hat{\bm z}\sin\eta)\nonumber\\
\hat{\bm m}&=&-\hat{\bm x}\sin\eta+\hat{\bm z}\cos\eta\nonumber\\
\hat{\bm n}&=&\hat{\bm y}\cos\phi-\sin\phi(\hat{\bm x}\cos\eta+\hat{\bm z}\sin\eta).
\label{e.lmnascr}\end{eqnarray}
Here we use cylindrical coordinates $r$, $\phi$ and $z$ centered at the hard core axis. The angle $\eta$ is a constant that describes the tilting of the $\hat{\bm l}(x,y)$ field out of the $x$-$y$ plane. Thus $\hat{\bm m}$ is a constant vector. In the order parameter (\ref{e.opmhca}) the amplitude of $\hat{\bm n}$ is modulated by function $f(r)$ so that it vanishes at the hard core axis, $f(0)=0$, and approaches unity outside of the hard core, $f(r\gg\xi)=1$.

Outside of the hard core, we can calculate the current using bulk A-phase hydrodynamics which works in the full temperature regime $0<T<T_c$ \cite{Cross1975}.   Since $\bm v_s=0$ in the form (\ref{e.lmnascr}), the current is given by \begin{eqnarray}
\bm j =C\bm\nabla\times\hat{\bm l}-C_0\hat{\bm l}(\hat{\bm l}\cdot\bm\nabla\times\hat{\bm l}).
\label{e.bapcr}\end{eqnarray}
Using (\ref{e.lmnascr}) gives currents in all cartesian directions, but only the  one in the $y$ direction does not average out in integration over $\phi$,
\begin{eqnarray}
j_y(x,y)=(C-C_0)\sin\eta\frac{\sin^2\phi}{r}=-\frac{4m_3K\Delta_A^2}{\hbar}\sin\eta\frac{\sin^2\phi}{r},
\label{e.jcdens}\end{eqnarray}
where the latter form gives the GL limit of the former. Note that $j_y(x,y)\propto y^2/(x^2+y^2)^{3/2}$ integrated over $x$ is independent of $y$. Thus the same total current has $x$ distribution whose width is given by $|y|$. That is, the peak value of current increases as the polar core is approched, in agreement with numerical calculations. Just at $y=0$ Eq.\ (\ref{e.jcdens}) would give a delta function, but in the hard core region we need to take into account the suppression factor $f(r)$ in the order parameter (\ref{e.opmhca}). A GL calculation gives
\begin{eqnarray}
j_y(x,0)= -\frac{4m_3K\Delta_A^2}{\hbar}\sin\eta\, f'(|x|).
\end{eqnarray}
We see that the $x$ distribution of the current is determined by the derivative of the suppression factor. It is noteworthy that the total current is again the same independently of the selected form of $f(r)$. We also note that the current is proportional to the tilt of the radial disgyration, $\sin\eta$. The model current (\ref{e.jcdens}) matches the numerical calculation in Fig.\ \ref{fig:xcurrents} excellently without fitting parameters.

We calculate the contribution of the hard polar core to the energy of the vortex. Previous works have roughly estimated the contribution to be on the order of a few percent~\cite{Vulovic1984}. The contribution is calculated from the final minimized texture by considering the region where the order parameter amplitude deviates more than $5\%$ from the bulk value, and comparing it to the energy of the whole vortex, including the flow contribution from outside the simulation box using Eq.~\eqref{eq:fkin2}. The outer cutoff is taken to be the intervortex distance $r_v = \sqrt{\hbar/2m_3\Omega}$ at $\Omega = 1\text{ rad/s}$, $r_v \approx 0.01\text{ cm}$. With this cutoff, the polar core contributes approximately $18.2\%$ of the total energy. We believe that a large portion of this energy originates from the mass flow through the core, which was not considered in previous works.

\section{Conclusion}\label{sec:conclusions}
We have calculated for the first time the detailed structure of the single-quantum vortex in the A phase of superfluid $^3$He. This allowed quantitative determination of the $\hat{\bm{l}}$ vector texture, the tilt of the radial disgyration in the hard core, the asymmetric shift of the hard and soft cores, and the energy contribution of the polar core to the whole structure. Similar to the double-quantum vortex in $^3$He-A, the SQV possesses the vortex topological invariant, related to circulation of the superflow, and also the skyrmion topological invariant, related to the distribution of the $\hat{\bm{l}}$ vector in the vortex core. Unlike in the DQV, the skyrmion invariant in the SQV is fractional, $N=1/2$. This forces the formation of a hard core and disgyration of the $\hat{\bm{l}}$ vector around it. This distribution of $\hat{\bm{l}}$ is found to have a dramatic effect on the supercurrent. The calculated superfluid current is highly asymmetric and reveals channeling through the hard core, which can be understood as the intrinsic angular momentum current, similar to the magnetization current in magnetic materials. The current is caused by the variation of the orbital momentum vector $\hat{\bm{l}}$ in a tilted disgyration around the SQV hard core.

Calculating experimental signatures of the SQV, such as the nuclear magnetic resonance response, and comparing them to the observations remains a task for future research. Especially interesting will be searching for consequences of the intrinsic orbital momentum currents and determination of the states and the energy spectrum of the core-bound fermions. It has been suggested that Weyl fermionic quasiparticles moving from the A phase with one orientation of the $\hat{\bm{l}}$ vector through the polar phase to the A phase with the opposite $\hat{\bm{l}}$ orientation experience effective tetrad gravity simulating transition to antispacetime \cite{Nissinen2018}. This situation is realized on a path through the polar core of the SQV. Taking into account strong flow through the core, which can entrain quasiparticles, one may wonder whether SQV in $^3$He-A can be used as a laboratory model of the Big Bang speculated as a transition from antispacetime \cite{Boyle2018}.
The quantitative calculation of the SQV structure allows also the completion of the theoretical phase diagram of vortices in $^3$He-A \cite{Karimaki1999}.

\backmatter


\bmhead{Acknowledgements}
The work was supported by Academy of Finland (present name: Research Council of Finland) project 332964. The calculations were performed using computer resources within the Aalto University School of Science “Science-IT” project.





\bibliography{sn-bibliography}

\end{document}